\documentclass[prd,onecolumn]{revtex4}
\usepackage{dcolumn}
\usepackage{multirow}
\usepackage{graphicx}
\usepackage{amssymb}
\usepackage{bm}
\usepackage{hyperref}
\usepackage{epstopdf}
\usepackage{color}
\usepackage{mathrsfs}
\usepackage{amsmath,amssymb,amsthm}
\usepackage{rotating}
\usepackage{sverb, longtable}
\usepackage{subfigure}

\usepackage[graphicx]{realboxes}
\usepackage{adjustbox}
\begin{document}

\title{Pantheon+ constraints on dark energy and modified gravity: An evidence of dynamical dark energy}
\author{Deng Wang}
\email{cstar@nao.cas.cn}
\affiliation{National Astronomical Observatories, Chinese Academy of Sciences, Beijing, 100012, China}
\begin{abstract}
We use the latest Type Ia supernovae sample Pantheon+ to explore new physics on cosmic scales. Specifically, in light of this new sample, we constrain the interacting dark energy and Hu-Sawicki $f(R)$ gravity models and employ the model-independent Gaussian processes to investigate whether there is an evidence of dark energy evolution. We find that Pantheon+ alone just gives weak constraints. However, a data combination of Pantheon+ with cosmic microwave background, baryon acoustic oscillations and cosmic chronometers gives a strong constraint on the interaction parameter $\epsilon=0.048\pm0.026$, which indicates that the energy may transfer from dark energy to dark matter in the dark sector of the universe at the $1.85\,\sigma$ confidence level. In the meantime, we obtain a very tight constraint on the deviation from general relativity $\log_{10} f_{R0}< -6.32$ at the $2\,\sigma$ confidence level. Interestingly, when combining Pantheon+ with cosmic chronometers and cosmic microwave background data, we find a quintessence-like dark energy signal beyond the $2\,\sigma$ confidence level in the redshift range $z\in[0.70,1.05]$. This implies the nature of dark energy may actually be dynamical.

\end{abstract}
\maketitle

\section{Introduction}
The standard cosmological model, $\Lambda$-cold dark matter ($\Lambda$CDM), has been well tested from cosmic scales to very small scales \cite{Will:2014kxa} by various kinds of observations such as Type Ia supernovae (SNe Ia) \cite{SupernovaSearchTeam:1998fmf,SupernovaCosmologyProject:1998vns}, baryon acoustic oscillations (BAO) \cite{Blake:2003rh,Seo:2003} and cosmic microwave background (CMB) \cite{Planck:2018vyg,WMAP:2003elm,Planck:2013pxb}. Although $\Lambda$CDM has displayed its success in explaining different cosmological phenomena, it still faces at least two intractable problems \cite{Weinberg:1988cp}: (i) why is the value of cosmological constant from current observations much smaller than that from the theoretical prediction? (ii) why are present densities of dark matter and dark energy (DE) of the same order? Besides, $\Lambda$CDM also suffers from small-scale crises such as the so-called ``Too-big-to-fail'' problem \cite{Boylan-Kolchin:2011lmk,Boylan-Kolchin:2011qkt}, i.e., the dark matter mass derived from stellar kinematics of stars within the half-light radius for the most massive observed satellites of the Milk Way (MW) galaxy is smaller than those of massive subhalos in the $\Lambda$CDM cosmological simulations of our MW galaxy \cite{Springel:2008cc}. To solve these problems and explore possible new physics, except the theoretical developments, we require new high-precision observations. Particularly, we need independent and powerful probes with better accuracies to give definite answers on some key puzzles. 

SNe Ia which help discover the cosmic acceleration are powerful geometrical distance indicators to investigate the background dynamics of the universe. During the past eight years, the first integrated SNe Ia sample is the ``Joint Light-curve Analysis'' (JLA) one constructed from the Supernova Legacy Survey (SNLS) and Sloan Digital Sky Survey (SDSS) in 2014, which consists of 740 data points covering the redshift range $z\in(0.01, \, 1.3)$ \cite{SDSS:2014iwm}. The second sample is the Pantheon one consisting of 1048 spectroscopically confirmed SNe Ia covering the redshift range $z\in(0.01, \, 2.3)$, which is constructed by combining 276 Pan-STARRS1 (PS1) SNe Ia with useful distance estimates of SNe Ia from SNLS, SDSS, low-z and Hubble space telescope (HST) observations \cite{Pan-STARRS1:2017jku}. Most recently, Refs.\cite{Scolnic:2021amr,Brout:2022vxf} report an updated Pantheon sample, called Pantheon+, which is made of 1701 light curves of 1550 spectroscopically confirmed SNe Ia coming from 18 different surveys. This larger sample is a significant increase of Pantheon, specially at low redshifts, and covers the redshift range $z\in[0.000122, \, 2.26137]$. In general, the direct usage of a new dataset is to explore possible new physics. As a consequence, in this study, first of all, we implement the first constraint on DE and modified gravity (MG) models by using this new SNe Ia sample. Furthermore, since the community is very interested in whether current observations can give an evidence of evolution of DE, we use Gaussian processes (GP), a model-independent statistical method, to reconstruct the equation of state (EoS) of DE by combining the Pantheon+ sample with other probes. We find that the dark energy may be dynamical over time beyond $2\,\sigma$ confidence level when using the data combination of Pantheon+, CMB and CC to reconstruct DE EoS.          

This study is organized as follows. In the next section, we introduce briefly the cosmological models to be constrained. In Section III, we describe the data and methodology to implement the constraints. In Section IV, we display the constraining results from observations. In Section V, we introduce briefly the GP methodology and data used. In Section VI, we exhibit our reconstructing results of DE EoS. The discussions and conclusions are presented in the final section.   

\section{Models}
Nowadays, there are two main approaches to resolve or even solve various cosmological puzzles, i.e., DE and MG. The former thinks that a kind of unseen matter component is responsible for the late-time cosmic acceleration, while the latter insists on that the gravity may be imperfect and the Lagrangian should be modified in order to satisfy the observational constraint. In the following context, we will introduce specifically interacting DE (IDE) and $f(R)$ gravity models and review briefly the corresponding basic formula. 

\subsection{IDE}
In the framework of general relativity (GR), the homogeneous and isotropic universe is described by the Friedmann-Robertson-Walker (FRW) metric
\begin{equation}
\mathrm{d}s^2=-\mathrm{d}t^2+a^2(t)\left[\frac{\mathrm{d}r^2}{1-\mathrm{K}r^2}+r^2\mathrm{d}\theta^2+r^2\mathrm{sin}^2\theta \mathrm{d}\phi^2\right],      \label{1}
\end{equation}
where $a(t)$ and $\mathrm{K}$ denote the scale factor at cosmic time $t$ and the Gaussian curvature of spacetime, respectively. Inserting Eq.(\ref{1}) into the Einstein field equation, we obtain the Friedmann equations governing the background dynamics of the universe as  
\begin{equation}
H^2=\frac{\sum\limits_{i}\rho_i}{3},     \label{2}
\end{equation}   
\begin{equation}
\frac{\ddot{a}}{a}=-\frac{\sum\limits_{i}(\rho_i+3p_i)}{6},     \label{3}
\end{equation}   

where $H\equiv\dot{a}/a$ is the Hubble parameter, dot is the derivative with respect to $t$, and $\rho_i$ and $p_i$ are energy densities and pressures of different components including radiation, baryons, DM and DE. Since focusing on the late-time universe, we neglect the radiation contribution to the cosmic evolution. Notice that we use the units $8\pi G=c=1$ throughout this study.

Since the nature of DM and DE is still not clear, in order to solve some cosmological tensions, it is natural to guess that there may be an interaction between DM and DE in the dark sector of the universe.  To realize the mechanism, one can build a decaying vacuum model \cite{Wang:2004cp} by considering a modified matter expansion rate $\epsilon$ during the evolution of matter $\rho_m=\rho_{m0}\,a^{-3+\epsilon}$, where $\rho_{m0}$ is present-day matter energy density. The Hubble parameter of this model reads as
\begin{equation}
H_{\mathrm{IDE}}(z)=H_0\sqrt{\frac{3\Omega_{m}}{3-\epsilon}(1+z)^{3-\epsilon}+1-\frac{3\Omega_{m}}{3-\epsilon}},   \label{4}
\end{equation}
where $\Omega_m$ is the matter fraction and the typical parameter $\epsilon<0$ indicates that the energy transfers from DM to DE and vice versa. Actually, one can easily find that the decaying vacuum model is a kind of IDE models by some algebraic derivations. Therefore, we refer to it as IDE in the following context. When $\epsilon=0$, this model naturally reduce to the $\Lambda$CDM case. 

Furthermore, we consider the linear perturbations of the IDE model in this analysis. In general, the scalar mode perturbation of FRW spacetime is expressed as \cite{Mukhanov:1992,Ma:1995,Malik:2009},
\begin{equation}
ds^2=-(1+2\Phi)dt^2+2a\partial_iBdtdx+a^2[(1-2\Psi)\delta_{ij}+2\partial_i\partial_jE]dx^idx^j,  \label{5}
\end{equation}
where $\Phi$ and $\Psi$ denote the linear gravitational potentials. Using the synchronous gauge $\Psi=\eta$, $\Phi=B=0$ and $E=-(h+6\eta)/2k^2$, the perturbed energy-momentum conservation equations of cosmic fluids are shown as \cite{Ma:1995}
\begin{equation}
\delta'=-3(\frac{\delta p}{\delta\rho}-\tilde{\omega})\mathcal{H}\delta-(1+\tilde{\omega})(\theta+\frac{h'}{2}), \label{6}
\end{equation}
\begin{equation}
\theta'=\frac{\delta p}{\delta\rho}\frac{k^2\delta}{1+\tilde{\omega}}+(3\tilde{\omega}-1)\mathcal{H}\theta-\frac{\tilde{\omega}'}{1+\tilde{\omega}'}\theta-k^2\delta, \label{7}
\end{equation} 
where $\tilde{\omega}$, $\mathcal{H}$, $\sigma$,  $\theta$ and $\delta$ are, respectively, the equation of state of cosmic fluids, conformal Hubble parameter, shear, velocity perturbation and density perturbation, and the prime denotes the derivative with respect to the conformal time. Subsequently, the perturbations of DE read as
\begin{equation}
\delta_{de}'=3\mathcal{H}(\omega_{de}-c_s^2)\left[\delta_{de}+3\mathcal{H}(1+\omega_{de})\frac{\theta_{de}}{k^2}\right]-3\mathcal{H}\omega'_{de}\frac{\theta_{de}}{k^2}-(1+\omega_{de})(\theta_{de}+\frac{h'}{2}),   \label{8}
\end{equation}
\begin{equation}
\theta_{de}'=\frac{c_s^2}{1+\omega_{de}}k^2\delta_{de}+(3c_s^2-1)\mathcal{H}\theta_{de}, \label{9}
\end{equation}
where $\omega_{de}$ and $c_s^2$ are the effective EoS of DE and the physical sound speed in the rest frame, respectively. In order to avoid the unphysical sound speed, we have set $c_s^2=1$. In the mean time, to calculate more smoothly, we also take $\sigma=0$ numerically. Note that the effective equation of state of IDE is $\omega_{de} =  -1+\frac{(1+z)^{3-\epsilon}-(1+z)^{3}}{\frac{3}{3-\epsilon}(1+z)^{3-\epsilon}-(1+z)^{3}+\frac{\tilde{\Omega_{\Lambda}}}{\Omega_{m}}}$ \cite{Wang:2004cp}, where $\tilde{\Omega_{\Lambda}}$ denotes the dimensionless ground state value of the vacuum.

\subsection{$f(R)$ gravity}
In principle, to construct a MG model, one should modify the geometrical term in Lagrangian. In this study, we will consider the simplest extension to GR, $f(R)$ gravity, where the modification is a function of Ricci scalar $R$. $f(R)$ gravity was firstly introduced by Buchdahl \cite{Buchdahl:1983zz} in 1970 and more details can be found in recent reviews \cite{DeFelice:2010aj,Sotiriou:2008rp}. Its action is shown as
 
\begin{equation}
S=\int d^4x\sqrt{-g}\left[\frac{f(R)}{2}+\mathcal{L}_m\right], \label{10}
\end{equation}
where $g$ and $\mathcal{L}_m$ are the trace of the metric and the matter Lagrangian, respectively. By varying Eq.(\ref{5}), we obtain the modified gravitational field equation as follows

\begin{equation}
f_RR_{\mu\nu}-\frac{f}{2}g_{\mu\nu}-\left[\nabla_\mu\nabla_\nu-g_{\mu\nu}\Box\right] f_R=T_{\mu\nu}, \label{11}
\end{equation}  

and after substituting the FRW metric into the above equation, we can obtain the Friedmann equation for $f(R)$ gravity as   
\begin{equation}
H^2+\frac{f}{6}-(H^2+H\frac{dH}{dN})f_R+H^2\frac{dR}{dN}f_{RR}=\frac{\rho_m}{3} ,              \label{12}
\end{equation} 
where the first derivative $f_R\equiv \mathrm{d}f/\mathrm{d}R$ iss an extra scalar degree of freedom, namely the so-called scalaron, the second derivative $f_{RR}\equiv \mathrm{d}f_R/\mathrm{d}R$ and e-folding number $N\equiv \mathrm{ln}\,a$.

We also consider the linear perturbations of $f(R)$ gravity. For sub-horizon modes ($k\gtrsim aH$) in the quasi-static approximation, the linear growth of matter density perturbations reads as \cite{Bean:2006up}
\begin{equation}
\frac{\mathrm{d}^2\delta}{\mathrm{d}a^2}+\left(\frac{1}{H}\frac{\mathrm{d}H}{\mathrm{d}a}+\frac{3}{a}\right)\frac{\mathrm{d}\delta}{\mathrm{d}a}-\frac{3\Omega_{m}H_0^2a^{-3}}{(1+f_R)H^2}\left(\frac{1-2X}{2-3X}\right)\frac{\delta}{a^2}=0, \label{13}
\end{equation}
where the function $X$ is expressed as 
\begin{equation}
X(k,a) = -\frac{2f_{RR}}{1+f_R}\left(\frac{k}{a}\right)^2. \label{14}
\end{equation}
It is worth noting that the function $X$ in Eq.(\ref{4}) gives a scale dependence of linear growth factor $\delta(k,a)$ in $f(R)$ gravity, when the growth factor is just a function of scale factor $a$ in GR. 

For the late-time universe, a successful $f(R)$ gravity model should explain the cosmic acceleration, pass the local gravity test and satisfy the stability conditions. To study the constraining power and cosmological implications of new Pantheon+ SNe Ia sample in the framework of $f(R)$ gravity, we take into account the viable Hu-Sawicki $f(R)$ model (hereafter HS model) \cite{Hu:2007nk} and it is depicted by 
\begin{equation}
f(R)=R-\frac{2\Lambda R^n}{R^n+\mu^{2n}}, \label{15}
\end{equation}  
where $n$ and $\mu$ are two free parameters characterizing this model. By taking $R\gg\mu^2$, the approximate $f(R)$ function can be expressed as 
\begin{equation}
f(R)=R-2\Lambda-\frac{f_{R0}}{n}\frac{R_0^{n+1}}{R^n}, \label{16}
\end{equation}
where $f_{R0}=-2\Lambda\mu^2/R_0^2$ and $R_0$ denotes the present-day value of Ricci scalar. Notice that when $f_{R0}=0$, HS $f(R)$ gravity reduces to $\Lambda$CDM.

\section{Data and methodology for constraints}
To study the constraining power of Pantheon+ sample and give joint constraints on representative DE and MG models, the datasets used in this analysis are listed below

$\bullet$ SNe Ia: SNe Ia, the so-called standard candle, is a powerful distance probe to study the background dynamics of the universe, particularly, the Hubble parameter and EoS of DE. We adopt the largest SNe Ia sample Pantheon+ to date, which consists of 1701 light curves of 1550 spectroscopically confirmed SNe Ia across 18 different surveys \cite{Scolnic:2021amr,Brout:2022vxf}. Pantheon+ has a significant increase relative to Pantheon at low redshifts and covers the redshift range $z\in[0.00122, \, 2.26137]$. 
This dataset is denoted as ``S''.

$\bullet$ CMB: Observations from the Planck satellite have extremely important meaning for cosmology and astrophysics. It has measured the matter components, topology and large scale structure of the universe. We employ the Planck-2018 CMB temperature and polarization data including the likelihoods of temperature at $30\leqslant \ell\leqslant 2500$ and the low-$\ell$ temperature and polarization likelihoods at $2\leqslant \ell\leqslant 29$, i.e., TTTEEE$+$lowE, and Planck-2018 CMB lensing data \cite{Planck:2018vyg}. We refer to this dataset as ``C''.

$\bullet$ BAO: The BAO probe which is hardly affected by uncertainties in the nonlinear evolution of matter density field and other systematic errors, is considered as a standard ruler to measure the background evolution of the universe. To break the parameter degeneracy from other observations, here we use four BAO measurements: the 6dFGS sample at effective redshift $z_{eff}=0.106$ \cite{Beutler:2011hx}, the SDSS-MGS one at $z_{eff}=0.15$ \cite{Ross:2014qpa}, and the BOSS DR12 dataset at three effective redshifts $z_{eff}=$ 0.38, 0.51 and 0.61 \cite{BOSS:2016wmc}. This dataset is identified as ``B''.

$\bullet$ Cosmic chronometers: We use the direct observations of cosmic expansion rate from cosmic chronometers (CC) as a complementary probe, which is independent of any cosmological model. This dataset is obtained by using the most massive and passively evolving galaxies based on the ``galaxy differential age'' method. In this study, we use 31 CC data points with systematic uncertainties \cite{Moresco:2020fbm} to carry out the constraints on the above two models. We call this dataset as ``H'' hereafter. 

To implement the constraints as done in our previous works \cite{Wang:2020dsc,Wang:2017klo,Wang:2017qtv}, at first, in synchronous gauge, we modify the publicly available Boltzmann code \texttt{CAMB} \cite{Lewis:1999bs} and incorporate the linear perturbations and background evolution of IDE and $f(R)$ models into it. Then, we also modify \texttt{CosmoMC} \cite{Lewis:2013hha} and take a standard Bayesian analysis via the Markov Chain Monte Carlo (MCMC) method to obtain the posterior distributions of free parameters. The last step is to analyze the MCMC chains with the public package \texttt{Getdist} \cite{Lewis:2019xzd}. Note that we use the Gelman-Rubin statistic $R-1=0.05$ as the convergence criterion of our MCMC analysis. For IDE, we choose the prior ranges for seven different parameters: $\Omega_bh^2 \in [0.005, 0.1]$, $\Omega_ch^2 \in [0.001, 0.99]$, $100\theta_{MC} \in [0.5, 10]$,  $\mathrm{ln}(10^{10}A_s) \in [2, 4]$, $n_s \in [0.8, 1.2]$, $\tau \in [0.01, 0.8]$ and $\epsilon \in [-3, 3]$. For $f(R)$, we employ the same priors except for we close the parameter $\epsilon$ and set $\log_{10} f_{R0} \in [-9, 1]$. 

In order to probe sufficiently the constraining power of Pantheon+, we will constrain two models using S alone and a data combination of C, B, H and S (hereafter CBHS), respectively. It is worth noting that, for simplicity, we just consider the $n=1$ case for HS $f(R)$ gravity in our constraints. The corresponding $\chi^2$ expressions for all the datasets can be found in Ref.\cite{Planck:2018vyg}.

\begin{figure}
	\centering
	\includegraphics[scale=0.55]{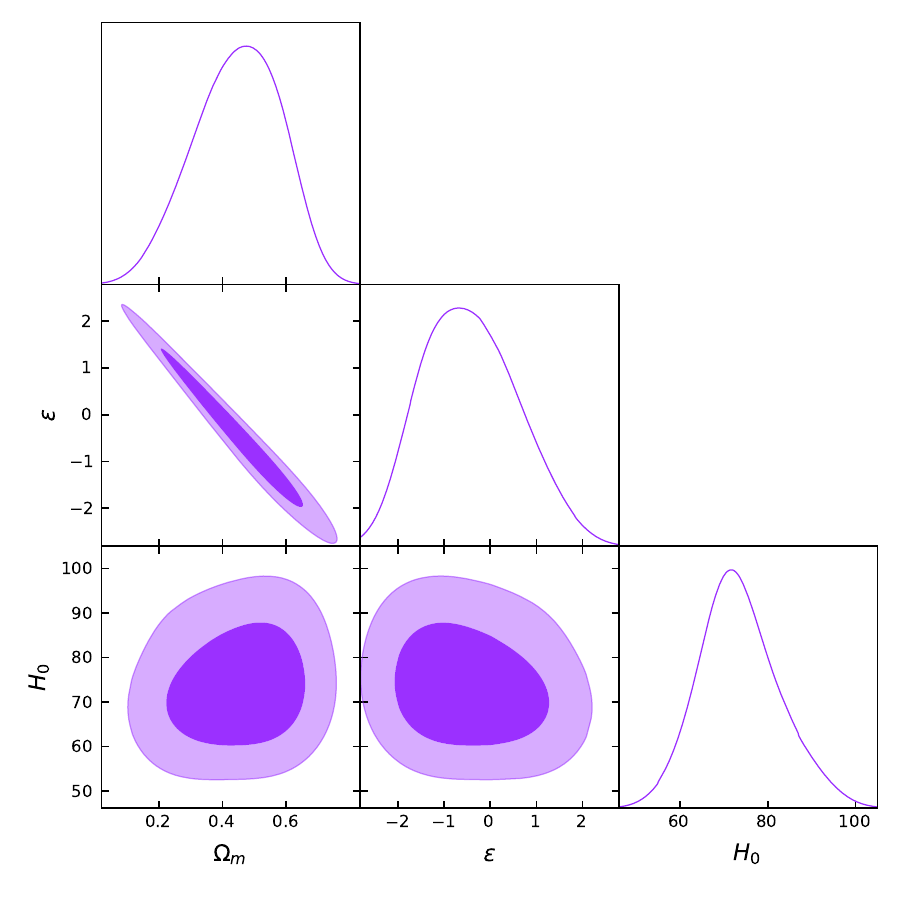}
	\includegraphics[scale=0.55]{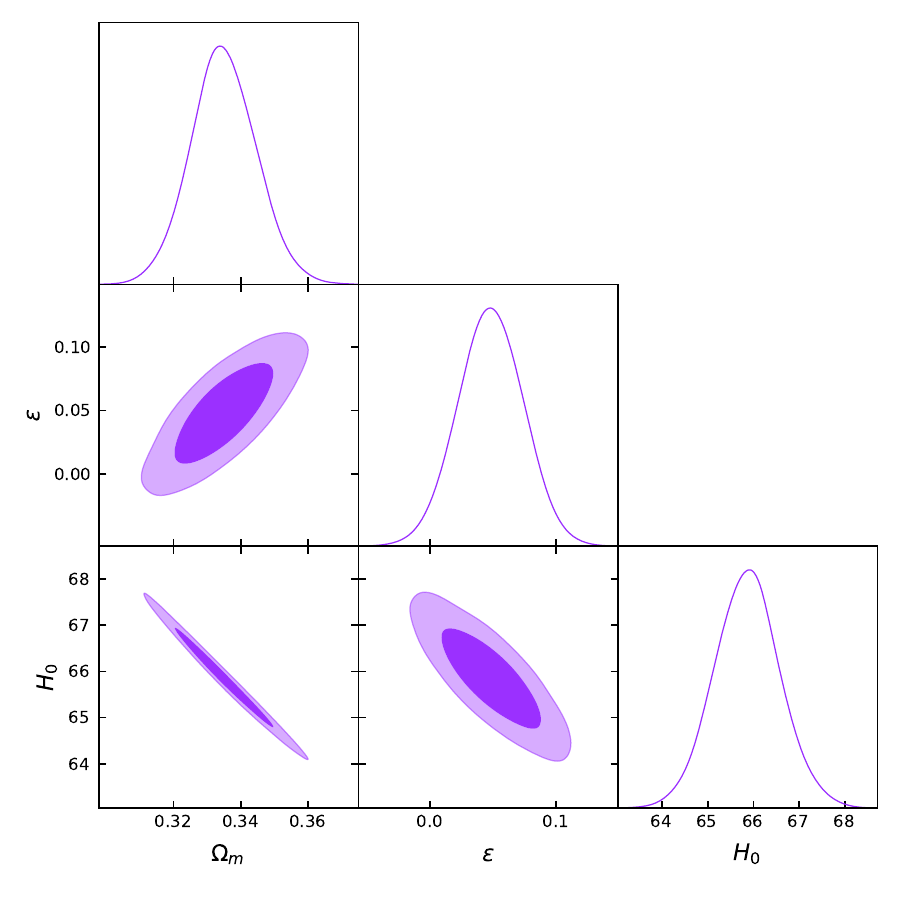}
	\caption{The marginalized $1\sigma$ and $2\sigma$ constraints on the IDE model by using the S (left) and CBHS (right) datasets, respectively. }\label{f1}
\end{figure}

\begin{figure}
	\centering
	\includegraphics[scale=0.55]{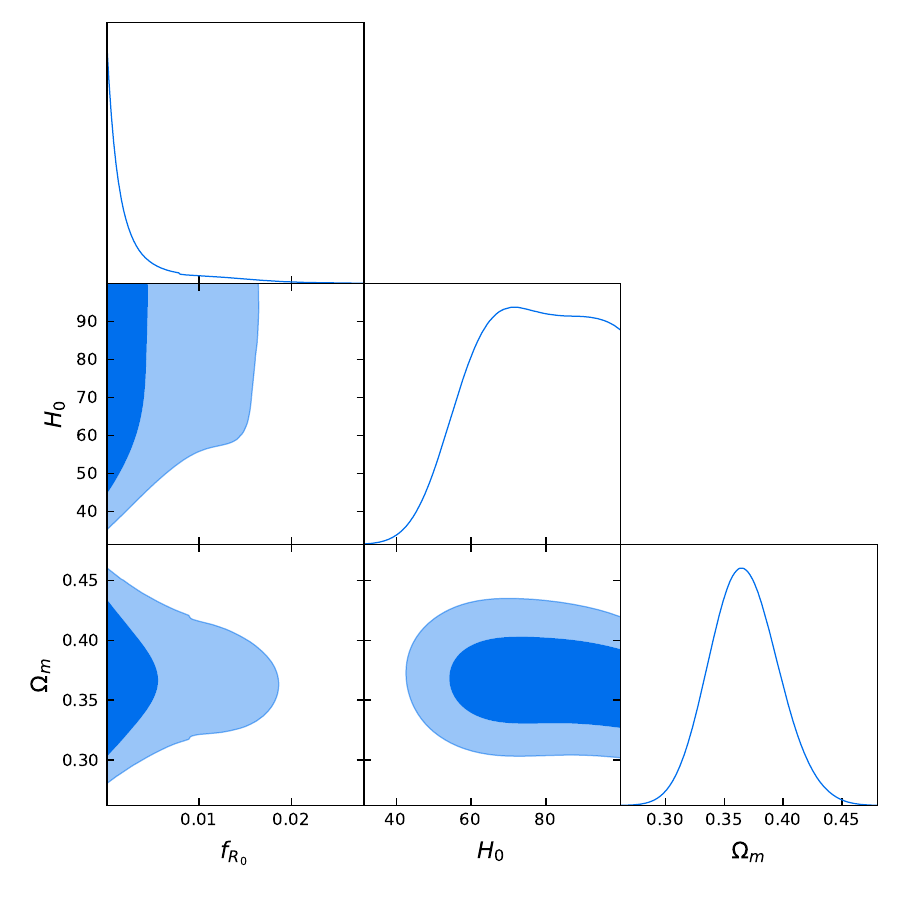}
	\includegraphics[scale=0.55]{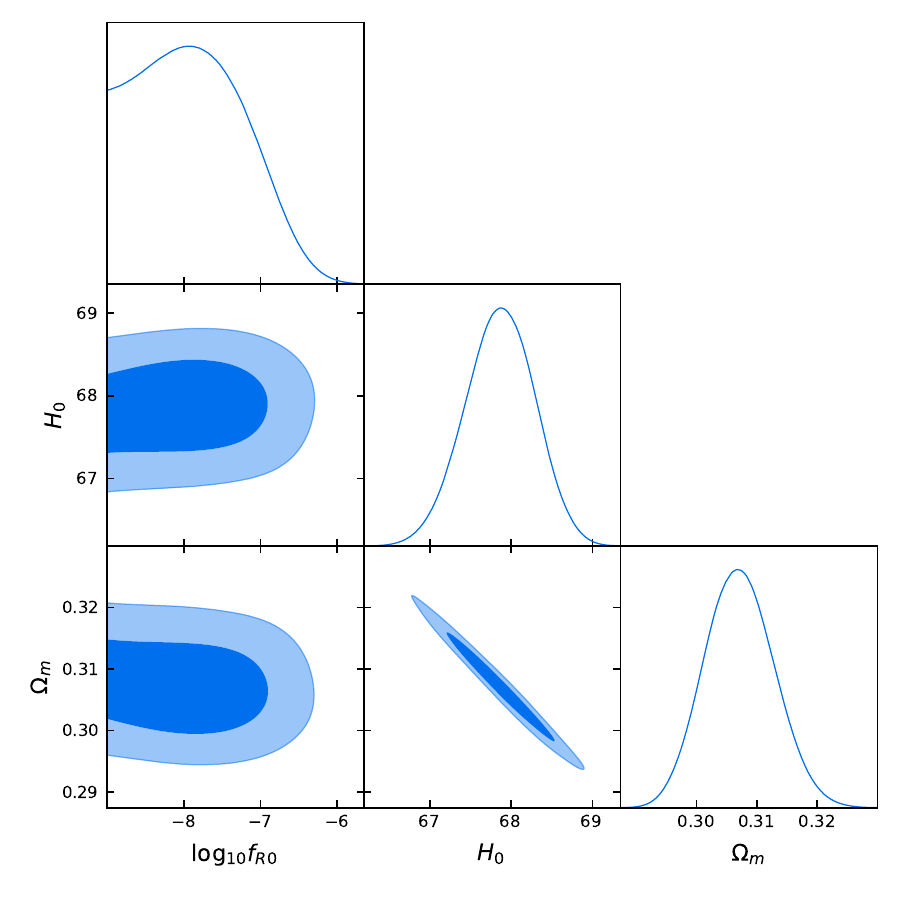}
	\caption{The marginalized $1\,\sigma$ and $2\,\sigma$ constraints on the HS $f(R)$ gravity by using the S (left) and CBHS (right) datasets, respectively.}\label{f2}
\end{figure}

\begin{figure}
	\centering
	\includegraphics[scale=0.5]{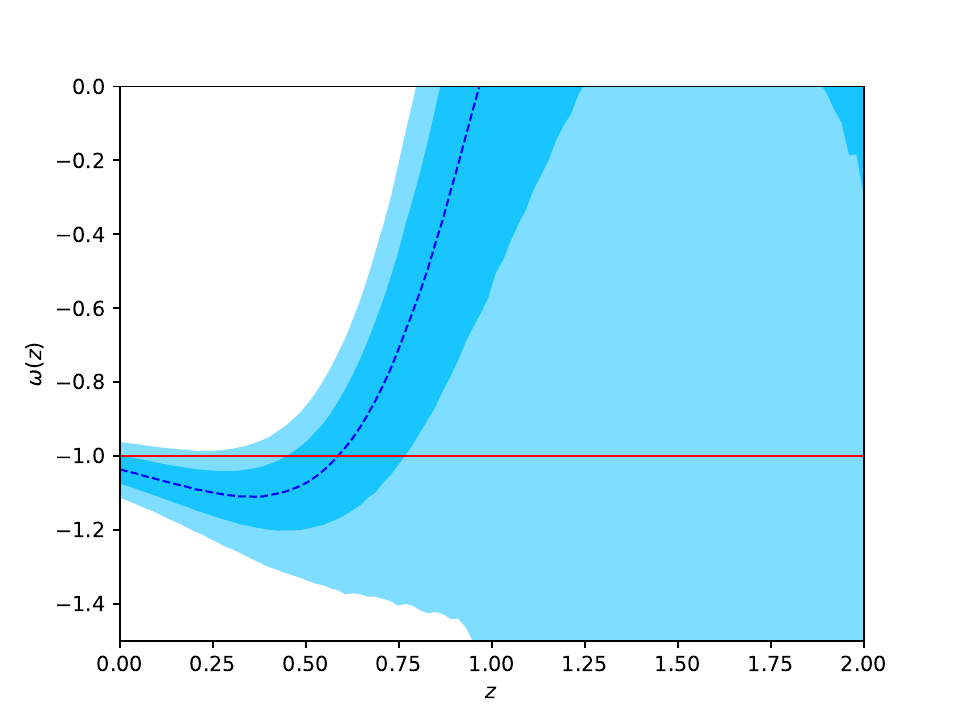}
	\includegraphics[scale=0.5]{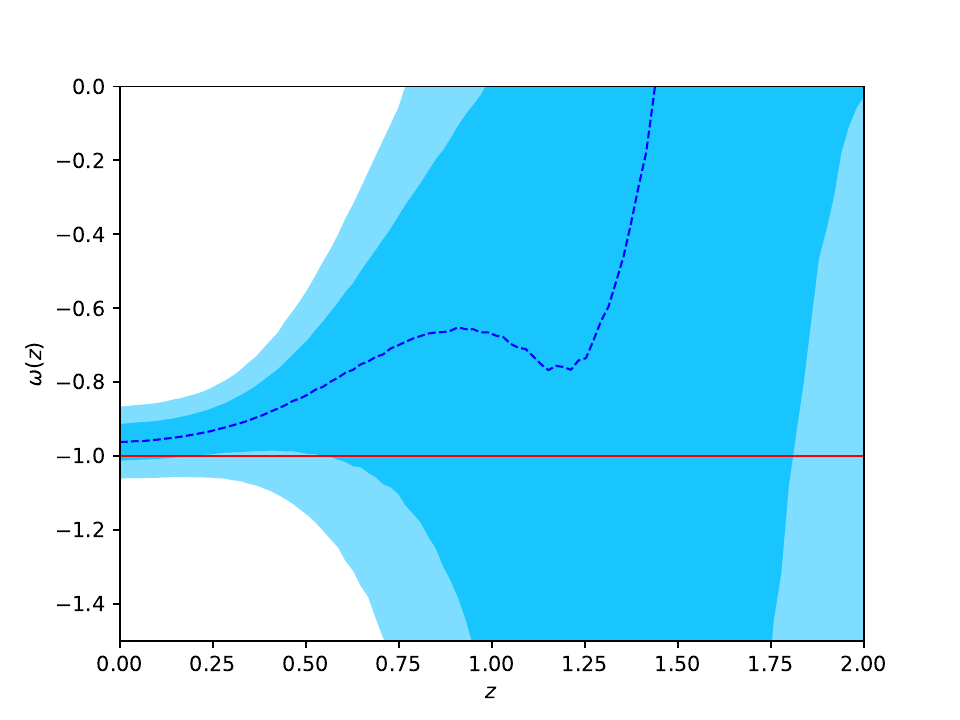}
	\caption{The GP reconstructions of EoS of DE by using the Pantheon (left) and Pantheon+ (right) datasets, respectively. The shaded regions denote the $1\,\sigma$ (dark) and $2\,\sigma$ (light) uncertainties of reconstructed DE EoS. The solid (read) and dashed (blue) lines are the $\Lambda$CDM and underlying DE EoS from data, respectively. We assume $H_0=73.04\pm1.04$ km s$^{-1}$ Mpc$^{-1}$ \cite{Riess:2021jrx}, $\Omega_m=0.315\pm0.007$ and $\Omega_k=0$ here.}\label{f3}
\end{figure}

\begin{figure}
	\centering
	\includegraphics[scale=0.5]{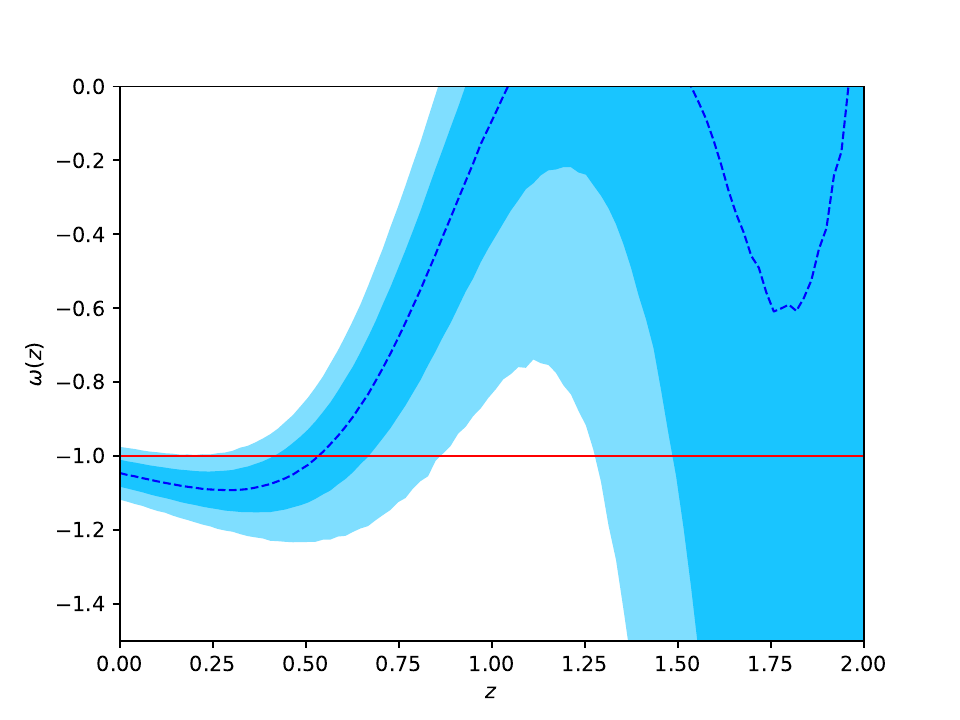}
	\includegraphics[scale=0.5]{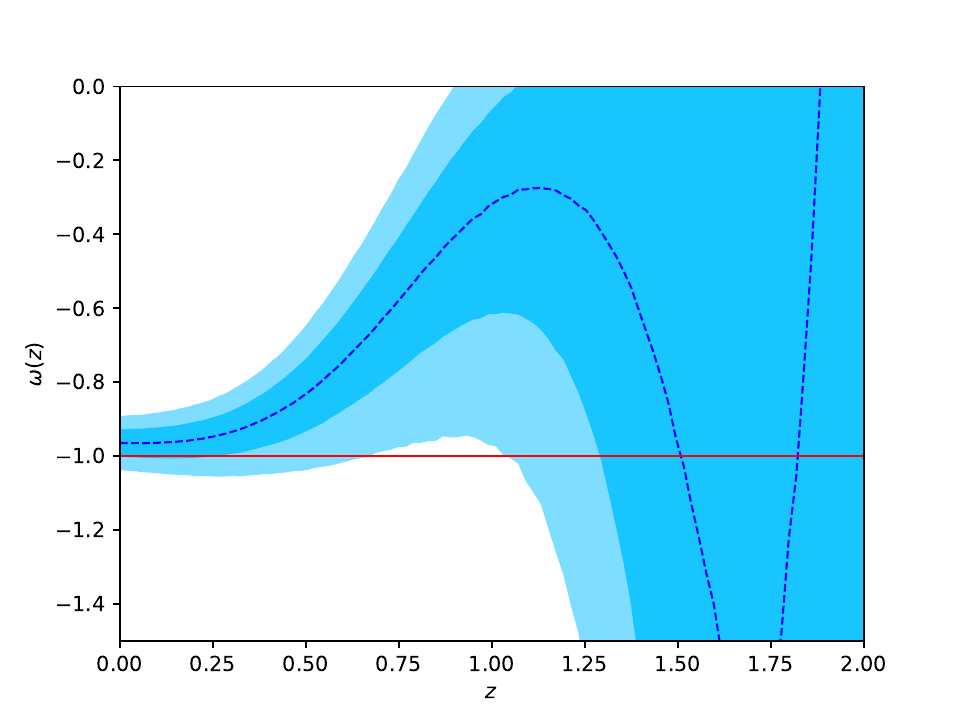}
	\caption{The GP reconstructions of EoS of DE by using the CMB+CC+Pantheon (left) and CMB+CC+Pantheon+ (right) datasets, respectively. The shaded regions denote the $1\,\sigma$ (dark) and $2\,\sigma$ (light) uncertainties of reconstructed DE EoS. The solid (read) and dashed (blue) lines are the $\Lambda$CDM and underlying DE EoS from data, respectively. We assume $H_0=73.04\pm1.04$ km s$^{-1}$ Mpc$^{-1}$, $\Omega_m=0.315\pm0.007$ and $\Omega_k=0$ here.}\label{f4}
\end{figure}

\begin{table*}[!t]
	\renewcommand\arraystretch{1.5}
	\caption{The mean values and uncertainties of free and derived parameters from marginalized constraints on the IDE and $f(R)$ gravity models are shown by using the  ``S'' and ``CBHS'' datasets, respectively.}
	\setlength{\tabcolsep}{7mm}{
	\begin{tabular} { l |c| c |c| c }
		\hline
		\hline
		
		Data                 & \multicolumn{2}{c}{S}      &   \multicolumn{2}{|c}{CBHS}                              \\
		\hline
		Model              &  IDE      & $f(R)$     &  IDE      & $f(R)$   \\
		\hline
		{\boldmath$\Omega_b h^2   $} & ---     & ---                &$0.02231\pm 0.00013        $    & $0.02233\pm 0.00014        $      \\
		
		{\boldmath$\Omega_c h^2   $} & ---    & ---                 & $0.12290\pm 0.00096        $  & $0.11815\pm 0.00094        $                                      \\
		
		{\boldmath$100\theta_{MC} $} & ---   & ---                  & $1.04085\pm 0.00031        $  & $1.04091\pm 0.00031        $                                        \\
		
		{\boldmath$\tau           $} & ---     & ---                & $0.063\pm 0.011            $   & $0.065\pm 0.012            $                                   \\
		
		{\boldmath${\rm{ln}}(10^{10} A_s)$} & ---  & ---             & $3.058^{+0.021}_{-0.022}   $   & $3.067\pm 0.022            $                                        \\
		
		{\boldmath$n_s            $} & ---   & ---                   & $0.9667\pm 0.0039          $  & $0.9672\pm 0.0040          $                                        \\
		
		{\boldmath$\epsilon$}  & $-0.44^{+1.51}_{-1.48}$    & ---  & $0.048\pm0.026$   & --- \\
		
		{\boldmath$\log_{10}  f_{R0}$}    & ---    & $< -1.89$ $(2\,\sigma)$   & ---    & $< -6.32$ $(2\,\sigma)$                                \\
		
		\hline
		
		{\boldmath$H_0                       $ }& $73.5^{+8.1}_{-10.0}                   $ & $> 52.5$ $(2\,\sigma)$    & $65.86\pm 0.72             $    & $67.78\pm 0.45$                        \\
		
		{\boldmath$\Omega_m                  $ }& $0.45^{+0.15}_{-0.13}   $ & $0.367\pm0.029$    & $0.3349\pm 0.0099          $  & $0.3058\pm 0.0060          $                                                    \\
		\hline
		\hline
	\end{tabular}
	\label{t1}}
\end{table*}
\section{Results for constraints}
The numerical analysis results of cosmological constraints on IDE and $f(R)$ models are shown in Fig.\ref{f1} and Fig.\ref{f2} and Tab.\ref{t1}. For the case of IDE, we obtain the modified matter expansion rate $\epsilon=-0.44^{+1.51}_{-1.48}$ and $0.048\pm0.026$ for S and CBHS datasets, respectively. One can easily find that although Pantheon+ is an updated version of Pantheon, it still provide a poor constraint on IDE. Very interestingly, when combining Pantheon+ with other datasets, we find a preference of positive $\epsilon$ at $1.85\,\sigma$ confidence level, which indicates that energy may actually transfer from DE to DM in the dark sector of the universe. For the case of $f(R)$, we find that Pantheon+ can just show a weak constraining power and gives $2\,\sigma$ upper bound $\log_{10} f_{R0}< -1.89$. But, if using the combined CBHS dataset, this limit is largely compressed to $\log_{10} f_{R0}< -6.32$, which is basically consistent with our previous constraint $\log_{10} f_{R0}< -6.75$, where we use a joint constraint from CBH, Pantheon and redshift distortions.

\section{Data and methodology for reconstructions}
To implement the GP reconstructions of DE EoS, we use the above-mentioned Pantheon+, CC and CMB datasets. We will use the CMB shift parameter $\mathcal{R}$ instead of the whole Planck dataset.    

In a FRW universe, the luminosity distance $D_L(z)$ is shown as
\begin{equation}
D_L(z)=\frac{c\,(1+z)}{H_0\sqrt{|\Omega_{k}|}}\mathrm{sinn}\left(\sqrt{|\Omega_{k}|}\int^{z}_{0}\frac{dz'}{E(z')}\right), \label{10}
\end{equation}
where the dimensionless Hubble parameter $E(z)\equiv H(z)/H_0$, the present-day cosmic curvature $\Omega_{k0}=-\mathrm{K}c^2/(H_0^2)$, and for $\mathrm{sinn}(x)= \mathrm{sin}(x), x, \mathrm{sinh}(x)$, $\mathrm{K}=1, \, 0, \, -1$ , which corresponds to a closed, flat and open universe, respectively. Subsequently, using the normalized comoving distance $D(z)=(H_0/c)(1+z)^{-1}D_L(z)$, the DE EoS reads as
\begin{equation}
\omega(z)=\frac{2(1+z)(1+\Omega_{k})D''-[(1+z)^2\Omega_{k}D'^2-3(1+\Omega_{k}D^2)+2(1+z)\Omega_{k}DD']D'}{3D'\{(1+z)^2[\Omega_{k}+(1+z)\Omega_{m}]D'^2-(1+\Omega_{k}D^2)\}}, \label{11}
\end{equation}
where the prime is the derivative with respect to the redshift $z$. It is worth noting that in our situation, the quantities $D, D', D''$ of Eq. (\ref{11}) can be obtained from the reconstructions, and the values of parameters $\Omega_{m}, \Omega_{k}$ will have impacts on the final reconstructions of DE EoS.

The GP can reconstruct a function from observations with no need of assuming a specific model or choosing a parameterized form for the underlying function \cite{Holsclaw:2010nb,Holsclaw:2010sk}.
Usually, the GP is a generalization of a Gaussian distribution.
At every reconstructing point $x$, the reconstructed function $f(x)$ is a Gaussian distribution with a mean value and Gaussian uncertainty. The important ingredient of GP is the covariance function $k(x,\tilde{x})$ which correlates the function $f(x)$ at different reconstructing points. In other words, the covariance function is solely determined by two hyperparameters $l$ and $\sigma_f$, which describe the coherent scale of the correlation in $x$-direction and change in the $y$-direction, respectively. As same as our previous analysis \cite{Wang:2017jdm,Wang:2017fcr,Wang:2019ufm}, we take the so-called Mat\'{e}rn ($\nu=9/2$) covariance function
\begin{equation}
k(x,\tilde{x})=\sigma_f^2\, \mathrm{exp}\left(-\frac{3|x-\tilde{x}|}{l}\right)\times\left[1+\frac{3|x-\tilde{x}|}{l}+\frac{27(x-\tilde{x})^2}{7l^2}+\frac{18|x-\tilde{x}|^3}{7l^3}+\frac{27(x-\tilde{x})^4}{35l^2}\right]. \label{12}
\end{equation}

We use a modified version of public package GaPP \cite{Seikel:2012uu} to implement our reconstruction. In our reconstruction, for Pantheon+ SNe Ia data points, we transform the theoretical distance modulus $m-M$ to $D$ as follows
\begin{equation}
m-M-25+5\lg(\frac{H_0}{c})=5\lg[(1+z)D]. \label{13}
\end{equation}
Clearly, we display the transformation relations between observational quantities of three probes and numerical quantities accepted by GaPP in the following manner (also see Refs.\cite{Wang:2017jdm,Wang:2017fcr,Wang:2019ufm}):
$$ \mathrm{Relations}
\Longrightarrow\left\{
\begin{aligned}
D & \Longrightarrow   m-M \\
D & \Longrightarrow  \mathcal{R}=\sqrt{\Omega_{m}}\int^{z_c}_0\frac{\mathrm{d}z'}{E(z')} \\
D' & \Longrightarrow  \frac{H_0}{H(z)}
\end{aligned}
\right\}
$$
where $z_c=1089.0$ is the redshift of recombination. Note that we consider a flat universe when implementing the reconstructions and use the CMB shift parameter $\mathcal{R}=1.7488\pm0.0074$  \cite{Planck:2015bue} in this study. It is reasonable to set the initial conditions $D(z=0)$ and $D'(z=0)=1$ when carrying out the GP reconstructions.

\section{Results for reconstructions}
To investigate whether dark energy is dynamical using the latest Pantheon+ sample, we reconstruct the DE EoS in the redshift range $z\in[0,2]$ by using Pantheon+ and CMB+CC+Pantheon+ datasets, respectively. Due to the new data release, we are also very interested in comparing the constraining power of Pantheon and Pantheon+. As a consequence, we also implement GP reconstructions using Pantheon and CMB+CC+Pantheon datasets, respectively. Our reconstructing results are shown in Fig.\ref{f3} and Fig.\ref{f4}. We find that the underlying DE EoSs from two SNe Ia samples is globally compatible with $\Lambda$CDM at $2\,\sigma$ confidence level, but there are some differences between them. $\Lambda$CDM is always close to the $2\,\sigma$ boundary of Pantheon reconstructed EoS when $z\in[0.05,0.45]$, is consistent with it when $z\in[0.45,0.75]$, and lies between its $1\,\sigma$ and $2\,\sigma$ bands when $z>0.75$ (see Fig.\ref{f3}). Interestingly, for Pantheon+, $\Lambda$CDM show completely different behaviors and always lie in the $1\,\sigma$ region of the underlying EoS. At first glance, this difference seems to indicate that Pantheon has a stronger constraining power than Pantheon+. However, concerning that Pantheon+ mainly increases the number of low-z SNe Ia relative to Pantheon, Pantheon+ will be more complete and this can let us obtain the correct and stable conclusion. It is noteworthy that Pantheon and Pantheon+ give the preferences of phantom-like $\omega(0)<-1$ and quintessence-like $\omega(0)>-1$ EoSs at $z=0$. This is very consistent with their cosmological constraints on $\Lambda$CDM [\cite{Brout:2022vxf}. Furthermore, to reduce statistical uncertainties on reconstructions, we include CC and CMB shift parameter observations in numerical analysis. Very interestingly, we observe that CMB+CC+Pantheon gives an obvious dynamical dark energy (DDE) signal beyond the $2\,\sigma$ confidence level in the range $z\in[0.85,1.30]$, while CMB+CC+Pantheon+ produces a relatively conservative quintessence-like DE signal in the range $z\in[0.70,1.05]$. One can also see that The inclusion of CC and CMB data can increase the constraining power to a large extent.               

\section{Discussions and conclusions}
The data release of Pantheon+ SNe Ia sample, which enhances largely the completeness of Pantheon sample, will help to explore the background evolution of the universe better. We make full use of this new sample to probe new physics. Specifically, we place constraints on IDE and HS $f(R)$ gravity models and use model-independent Gaussian processes to probe whether there is a hint of dark energy evolution. 

We find that Pantheon+ as an independent probe gives relatively weak constraints on IDE and $f(R)$ models. After combining it with CMB, BAO and CC, we obtain the constraint on the interaction parameter $\epsilon=0.048\pm0.026$, which implies that the energy may transfer from DE to DM in the dark sector of the universe at the $1.85\,\sigma$ confidence level. This constraint may produce more DM, lead to more collapsed structures in the late universe, and consequently worsen the ``Too-big-to-fail'' problem. In the meanwhile, we give a strict $2\,\sigma$ upper bound on the deviation from GR $\log_{10} f_{R0}< -6.32$. It is easy to conclude that Pantheon+ occupies a smaller weight than CMB or BAO in the combined constraint due to its weak constraining power. We still need a more complete and higher precision SNe Ia sample to compete with CMB or BAO in exploring the background dynamics of the universe.    

Due to low-z SNe Ia increase in Pantheon+ relative to Pantheon, we obtain a relatively stable reconstruction of DE EoS using Pantheon+. Using Pantheon+ or Pantheon, we find that the reconstructed DE EoS is always consistent with the prediction of $\Lambda$CDM within the $2\,\sigma$ confidence level. However, when combining Pantheon+ with CMB and CC, we find a quintessence-like DE signal beyond the $2\,\sigma$ confidence level in the range $z\in[0.70,1.05]$. This conclusion requires more reliable independent probes to implement a cross check. 

One may question that why we use two different methods, i.e., model-dependent and model-independent methods, to carry out the constraints. The answer is that, when we have a new sample of background data, in order to know whether this sample can give a global signal of new physics, the best way is implementing a model-dependent constraint. However, if one wants to know the evolution properties of new physics or a key cosmological quantity, the best approach is implementing a model-independent constraint and we use GP in this analysis. Both methods are complementary and their simultaneous applications in the analysis should be a complete paradigm to study the specific new physics.

Note that although current SNe Ia sample is still incomplete, the data combination of Pantheon+ (or Pantheon), CMB and CC still gives strong evidence of DDE, especially in the redshift range $z\in[0.70,1.30]$. This indicates that DDE may actually exist.
    
\section{Acknowledgments}
DW is supported by the National Science Foundation of China under Grants No.11988101 and No.11851301.

\end{document}